Red deposit (cinnabar) is evidenced on the internal side of a Neolithic pottery found on the right bank of the River Danube. This finding is related to an early processing of mercury ore.

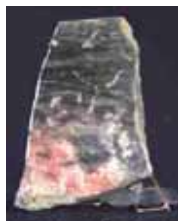

U.B. Mioc, Ph. Colomban, G. Sagon, M. Stojanović and A. Rosić

*Ochre decor and Cinnabar Residues in Neolithic Pottery from Vinča, Serbia*

# Ochre decor and Cinnabar Residues in Neolithic Pottery from Vinča, Serbia

U.B. Mioč[1], Ph. Colomban[2*], G. Sagon[2], M. Stojanović[3] and A.Rosić[4]

[1]Faculty of Physical Chemistry, University of Belgrade, P.O. Box 550, 11001 Belgrade, Serbia and Montenegro

[2]Laboratory for Dynamics, Interaction and Reactivity (LADIR), UMR7075 CNRS & University Pierre & Marie Curie, 2 rue Henry-Dunant, 94320 Thiais, France.

[3]National Museum Belgrade, Diana Center for conservation, Cara Urosa 20, 11001 Belgrade, Serbia and Montenegro

[4]Faculty of Mining and Geology, Department for Crystallography, University of Belgrade, Đušina 7, 11001, Belgrade, Serbia and Montenegro

**Abstract**

The prehistoric site of Vinča, on the right bank of the River Danube, the territory of the City of Belgrade, first excavated by Dr Miloje Vasić (1931-1934) provides Neolithic pottery dating back to 5200 – 4200 B.C. Shards excavated in 1998 (Serbian Academy of Sciences and Arts) have a yellow coating on the external (convex) and red deposit on the internal (concave) side. Raman, IR and X-ray identification prove that yellow-to-red decor deposited on the external faces of pottery is made of ochre, a mixture of hematite, quartz and phyllosilicates. Red deposit, found on some internal surfaces of a pottery, consists of cinnabar (HgS) with some quartz and phyllosilicates. This indicates that cinnabar was not used for decor but for some other purposes, preparations made in ceramic utensils. A comparison is made with the mercury ore from a Šuplja Stena mine located ~ 20 km from the Vinča village.

# 1. INTRODUCTION

Excavations at the prehistoric site of the Vinča started at the beginning of the 20th century by Dr Miloje Vasić (1931-1934) [1,2] on the territory of the City of Belgrade on the right bank of the River Danube (Fig.1). The importance of this site for archaeology of the Neolithic Era in Europe is extraordinary, because of the 10 meters thick cultural layer, which provides informations on all the events in the region of Southeast Europe between 5200 and 4200 B.C. Famous artifacts of the Vinca culture are small statues (height 10 cm) as shown in Plate 1. Another important finding of the Vinča culture is the use of cinnabar, malachite and kilns suggesting the existence of well-developed technology of pigments and early copper metallurgy [3].

Dr M.Vasić claims that the inhabitants of Vinča had used the cinnabar ore as a red pigment, and processed it to get mercury. Cinnabar was found in all cultural layers of the site and it was concluded that its origin was from the mine of Šuplja Stena under mount Avala, which is some 20 km away from Vinča [1,4]. New excavations were carried out (1998-2002) under the auspices of the Serbian Academy of Sciences and Arts. An interdisciplinary approach of this investigation was directed at gaining an insight into the way of life of the ancient inhabitants of Vinča and their technological development.

Pottery shards under study are from the layer, representing the late Neolithic civilization and are characteristic for some changes in decorations. The technology of making pottery is standard in Neolithic period. This pottery exhibits characteristic black color resulting from a firing under reduced atmosphere. However it has been found that low-temperature firing in reduced atmosphere leads to a better densification because FeO acts as a fluxing agent [4]. Shards decorated with "coloured earth" from the excavation site and a piece of cinnabar ore from the Šuplja Stena mine, mount Avala, are presented in Plate 1. There are evidences of its use even in the Neolithic age. In the Roman period the interest in this mine decreased because the mines Almadena in Spain and Idrija in Slovenia, conquered by the Romans were much richer mines of this ore.

# 2. EXPERIMENTAL

## 2.1. Samples

Plate 1 compares internal (concave) and external (convex) sides of the excavated shards. The largest 13p/78 sample size is 8x5 cm$^2$, thickness ~4-6 mm. The smallest c-137 sample size is 2.5x2 cm$^2$, thickness ~3mm. The pottery is free of any glaze but has rather smooth surfaces (rough polishing). The body colour is a grey-to-black. Red and yellow deposits are observed on external (samples 13p/78, c-59, c-137, c-371) or internal (13p/78, c-59, c-155, c-417) sides. For the sample c-59, yellow deposit is observed on both sides. Red colour is observed on samples c-137, c-155, c-161, c-371, and c-417. For samples 13p/78 yellow deposit is located in a scratch of the artefact. Intentional deposit can be questioned while the contamination from the soil is possible. In sample c-59 however the yellow deposit covers the upper part of the ceramic vase nicely and intentional deposit is likely. A well-defined distribution of the red deposit is observed on the c-155 sample (internal side). Heterogeneous distribution is observed on other samples.

## 2.2. Techniques

All previously mentioned samples were investigated using several spectroscopic methods: micro-Raman and infrared (IR) spectroscopy as well as X-ray powder diffraction (XRPD).

A multichannel notch-filtered INFINITY spectrograph (Jobin-Yvon–Horiba, France) equipped with a Peltier cooled CCD matrix was used to record Raman spectra between 150 and 3500 cm$^{-1}$, using 532 (or 632.8) nm exciting lines (0.01-5 mW power). Illumination and collection of the scattered light were made through an Olympus confocal microscope (long focus Olympus x50 objective, total magnification x 500). An "XY" spectrograph (Dilor, France) equipped with a double monochromator filter and a back-illuminated, liquid nitrogen-cooled, 2000 x 256 pixels CCD detector (Spex, Jobin-Yvon–Horiba Company) was also used and this allowed recording down to ca. 10 cm$^{-1}$ with 647.1 nm exciting line (slit = 80 μm). The 457 nm excitation (0.1 to 5 mW) was used to obtain a large spectral window. 568.2 and 514.5 nm exciting radiations were also used. Various objectives from MSPlan, Japan (numerical aperture = 0.80; magnification = 10, 50 and 100) were used; the total magnifications were 100, 500 and 1000; the confocal hole aperture was ~ 100 μm. Rare spikes (cosmic ray) and some plasma lines were removed from some spectra.

Infrared spectra of pigment, ore and "coloured earth" were recorded on a Perkin Elmer 983 G model spectrophotometer in the region from 4000 to 180 cm$^{-1}$. Samples were prepared as pellets of small amounts of ground pottery powder in KBr or CsI depending on the spectral region.

The X-ray powder diffraction (XRPD) patterns were recorded in the same conditions on a Philips PW-1710 automated diffractometer using a Cu-tube ($\lambda_{CuK\alpha}$ = 1,54178 Å). The instrument was equipped with a diffracted beam curved graphite monochromator. Diffraction data were collected in the ranges of 2θ Bragg angles, 5-50°, at 0.02° steps.

## 3. RESULTS

### 3.1. Micro-Raman and infrared analyses

The pigments on samples of pottery: c-59; c-100; c-137; c-155; c-371 and c-417 were analysed by micro-Raman spectroscopy. Results of the examination of pottery samples, Fig. 2, show two kinds of Raman signatures of red pigments i) a strong band at 254 and doublets at 282-290 and 343-352 $cm^{-1}$ assigned clearly to cinnabar, HgS, [5,6] and bands at ~225, 292, 410, 505 and 1315 $cm^{-1}$ assigned to hematite ($\alpha$-$Fe_2O_3$) ; small contributions at ~620-660 $cm^{-1}$ indicate the presence of more reduced iron oxide (magnetite, $Fe_3O_4$) [7-11]). All our samples can be divided into three groups: the sample c-417 where cinnabar was identified as a single cinnabar pigment, samples c-100, c-137, c-155 and c-371 with red pigment, where the strongest band in the Raman spectra is at 290 $cm^{-1}$ and the other characteristic bands of hematite.

Some other interesting Raman signatures are also observed. In the spectrum of c-59 sample a narrow band at ~140 $cm^{-1}$, and broad ones at 410 and 560 $cm^{-1}$ are assigned to $TiO_2$ (rutile, or a mixture of rutile and anatase [8-10]). In the case of sample c-137, the spectra of black spots indicate the presence of carbon; this is consistent with the "black" colour of the samples and the observation of magnetite signature in some points. A narrow peak at 1085 $cm^{-1}$ characteristic of calcite is also observed.

On Fig. 3 the IR spectra of a ground cinnabar single crystal (A), the ore from the Šuplja Stena (B), and that of the "coloured earth" (C) from the Vinča archaeological site are compared. In the spectrum (A) the presence of cinnabar is confirmed on the base of characteristic bands at 345, 280 cm$^{-1}$. The presence of cinnabar is evident in the ore, spectrum (B), and the traces of silicate minerals (bands at: ~1028, ~538 and ~470 cm$^{-1}$) and quartz (bands at: doublet at ~ 794 and 778, 465 and 207 cm$^{-1}$) were also found.

The spectrum of "yellow-to-red coloured earth" (C) is much more complex and structured bands about 1000, 540, 460, 350 cm$^{-1}$ are evident. Characteristic bands are assigned to: quartz (794, 787, 465, 207 cm$^{-1}$), silicate minerals (~1028, ~540 and ~470 cm$^{-1}$), calcite (~ 1080, ~ 1420 cm$^{-1}$), α-hematite (~1160, 630 and 530 cm$^{-1}$) . It could be supposed that red colour of "coloured earth" has origin in iron oxides α-hematite or limonite. Results of IR analysis were confirmed by XRD analysis.

**3.2. XRPD analysis**

XRD analysis of the ore from the Šuplja Stena mine shows the presence of cinnabar mineral (HgS) more or less mixed with quartz and other silicates minerals.

A sample of "coloured earth" was analysed too. The samples contain quartz (d = 4.24, 3.35, 2.46, 2.28, 2.13 Å) as the main constituent and phyllosilicates: mica (d = 9.9, 4.95, 4.44, 3.32, 2.56 Å), chlorite (d = 13.9, 7.03, 4.69, 3.52, 2.56 Å), and smectite (d = 15.2, 4.46, 2.56 Å). Small amounts of feldspar from plagioclase group (d = 4.02, 3.77, 3.20, 3.18 Å) and carbonate minerals: dolomite (d = 2.89 Å) and calcite d = 3.02 Å) are also present. Broad peaks are ascribed to hematite (α-$Fe_2O_3$) (d = 2.68, 2.51, 2.20, 1.85 Å).

Pigment on the sample c-161 extracted with a cutter was analysed. XRD patterns reveal certain amounts of the following minerals in descending order: quartz (d = 4.25, 3.35, 2.46, 2.28, 2.23, 2.13 Å), phyllosilicates: mica (d = 10.0, 5.0, 4.48, 2.56 Å), kaolinite (d = 7.15, 3.58 Å) and some smectite (17, 4.5, 2.6 Å), feldspar (d = 3.18 Å) a very small amount, as well as carbonate minerals: calcite (3.04 Å) and dolomite (d = 2.89 Å). Traces of hematite (d = 2.70, 2.52, 2.20 Å) are present. These results indicate that a small amount of ceramic base was also taken during sample extraction.

JCDD reference standard cards [12], used for identification of the mentioned minerals are: 05-0490 (quartz); 34-0175 (mica); 13-0003 (chlorite); 29-1497 (smectite); 10-0359 (feldspar); 36-0426 (dolomite); 47-1743 (calcite); 06-0502 (hematite) and 29-1488 (kaolinite).

## 4. DISCUSSION

Results of XRPD and vibration analyses of the ceramic body are in agreement: quartz and phyllosilicates are the dominant components. Carbonates such as calcite have been confirmed with the methods of vibration spectroscopy, too. This indicates that the firing temperature is lower than the temperature of complete transformation of calcite and phyllosilicates (~<900°C)

Hematite ($\alpha$-$Fe_2O_3$) present in both, pigments and coloured earth, shows that there is possibility of finding the raw material for these pigments in the surrounding due to close similarity in their mineral composition. The presence of chemically-combined water in some minerals such as smectite group and kaolinite, found in pigments and the "coloured earth" by XRPD, indicates that the pigment had been added to pottery after its firing.

Based on the fact that cinnabar has melting at $344^0C$ and becomes black above this temperature it could be taken for certain that the pigment had not been fired. Pieces of cinnabar were also found at all levels of the Vinča layers as reported by Dr Miloje Vasić the first researcher of this locality [1]. The fact that the cinnabar deposit was found in ceramic utensils confirms that cinnabar was collected, separated from the other ore phases and used for special purpose but not only as a ceramic pigment.

## 5. CONCLUSION

The analyses of pottery fragments from the Vinča archeological site have shown that the inhabitants of this area used two kinds of red pigments cinnabar and red ochre from the surroundings. From the analysis of pigments, ceramics body, "coloured earth", and ore from the Šuplja Stena mine – mount Avala we are pretty certain that these pigments came from the territory of central zone of Vinča influence. The technique of painting with red hematite pigment is characteristic of the late Vinča period and represents the change in the repertoire of pottery ornamentation. Cinnabar was prepared or stored in ceramic utensils for special purposes.

**ACKNOWLEDGMENTS**

Special thanks are due to Mr N. Tasić, the chief of Department of Methodology in Archaeology for cooperation and suggestions, Z. Nedić, P. Dakić and B. Petrović, for their technical support.

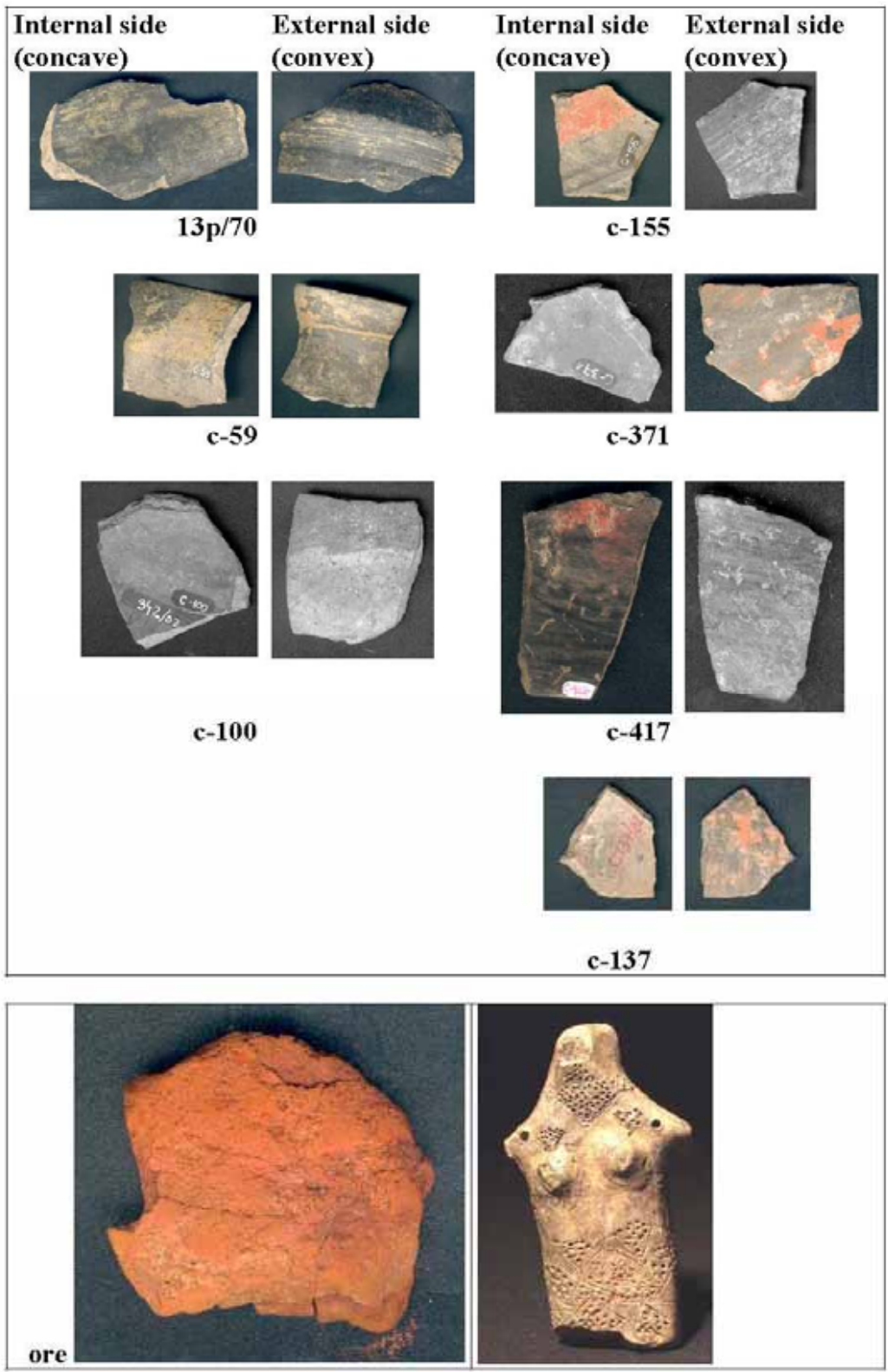

Plate 1. Shards excavated at the Vinča site (13p/78, c-59, c-100, c-137, c-155, c-371, note the red deposit on c-417 internal side) and a sample of red "coloured earth" ore from Šuplja Stena mine – mount Avala. An example of typical statue is shown.

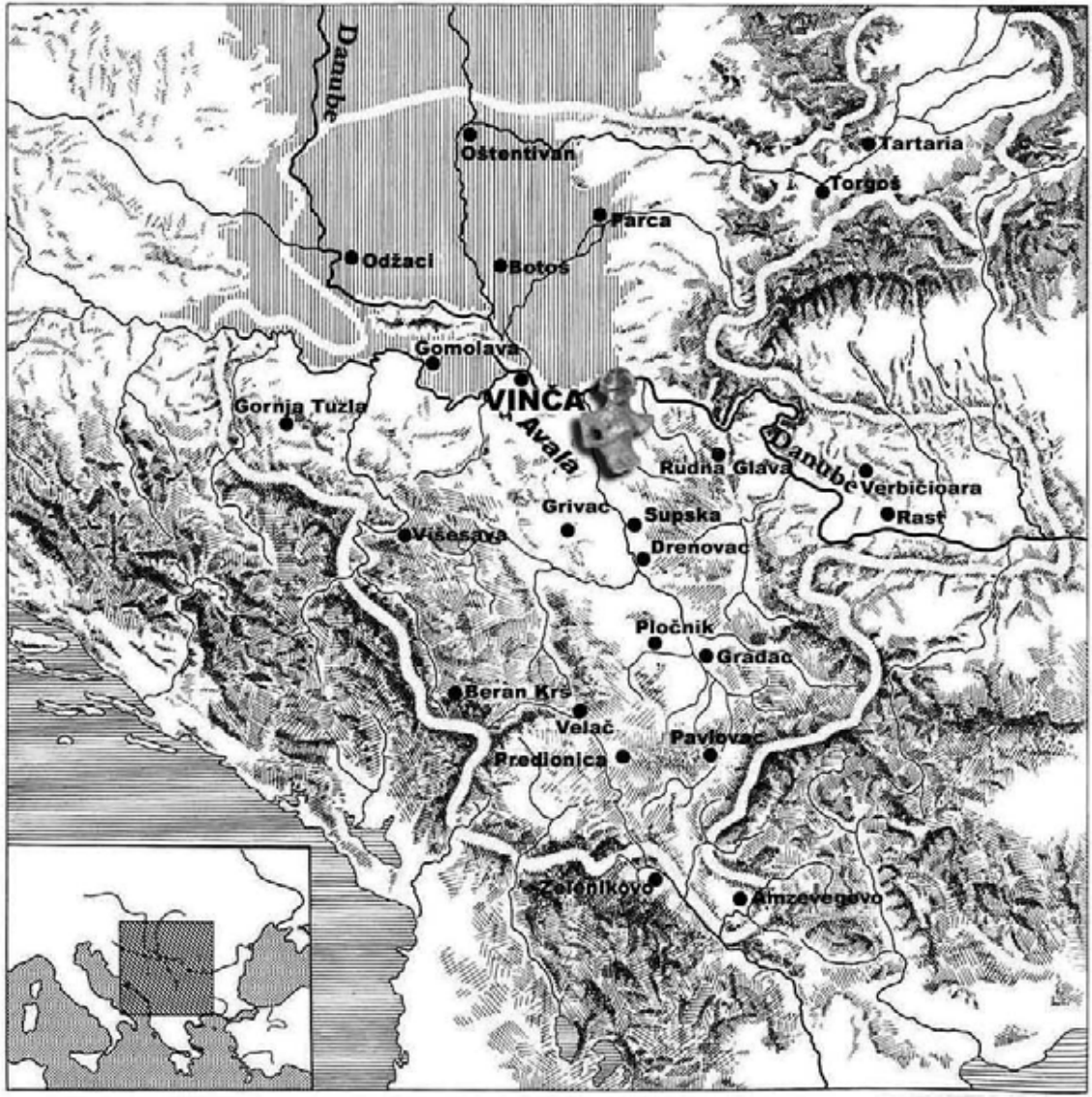

Figure 1. Territory of Vinča culture influence; see the location of the Vinča village on the the right bank of river Danube. Cinnabar ore was found under the Avala mountain, ~20 km from Vinča.

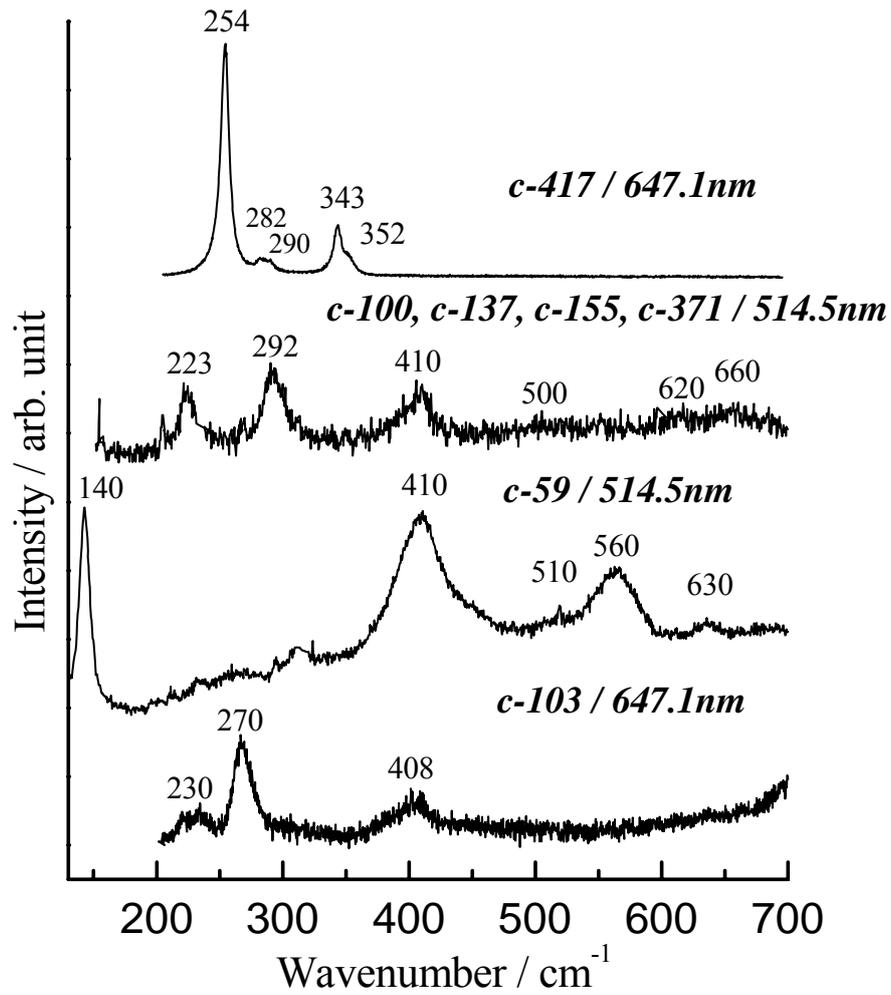

Figure 2. Micro-Raman spectra of red deposits: c-417 sample (647.1 nm, 0.140 mW) and c-137 sample (514.5nm, 0.230mW); similar spectra are observed for C-100, c-155 and c-371 samples. Examples of other observed Raman signatures are given.

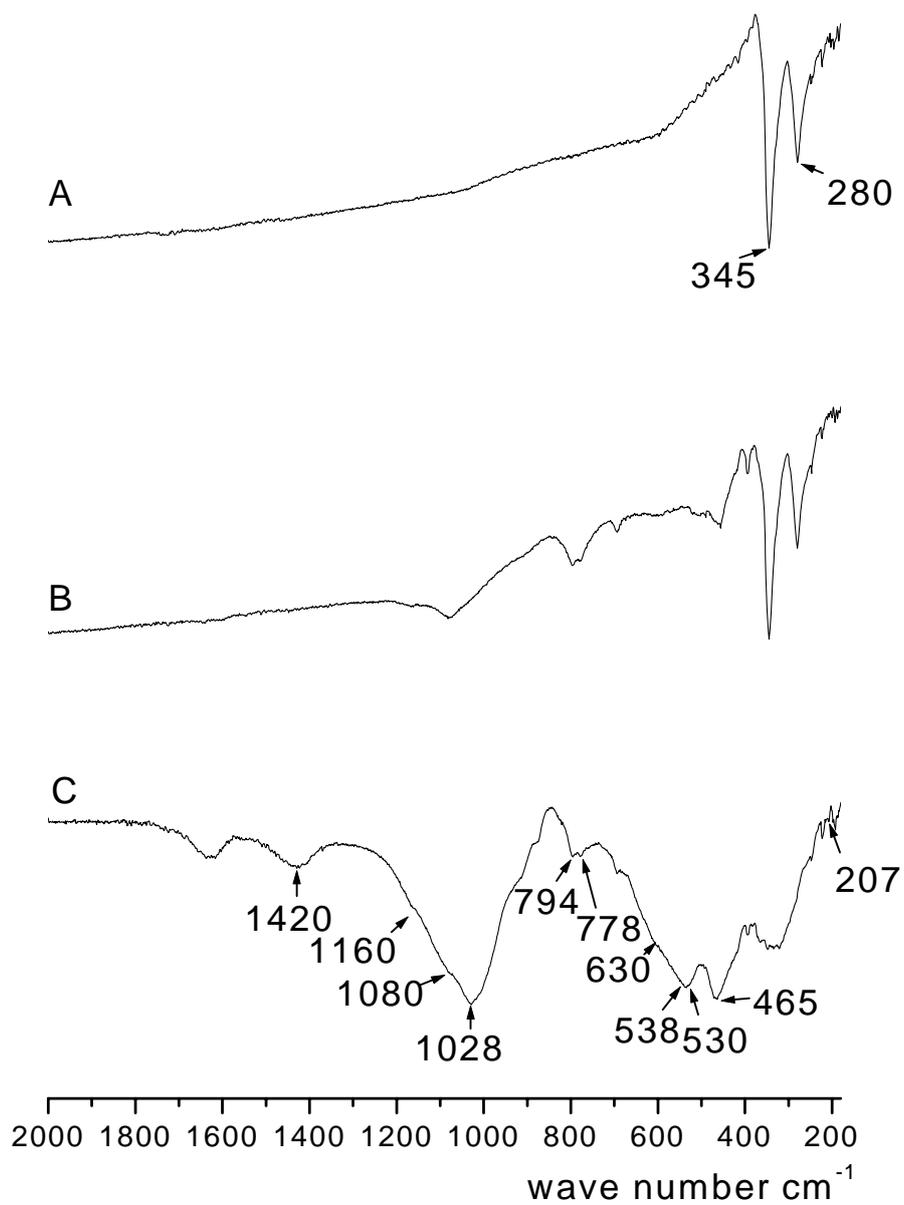

Figure 3. Infrared spectra of single crystal of cinnabar (A), cinnabar ore (B) from mount Avala; "coloured earth" arhelogical site the Vinča (C).